\documentclass[aps,prd,preprint,showpacs]{revtex4}

\usepackage{longtable}
\usepackage{textcomp}
\usepackage{multirow}

\begin{document}
\title{Intrinsic charm content of the nucleon and charmness-nucleon sigma term}

\author{Shaorong Duan$^{1}$}

\author{C. S. An$^{1}$}\email{ancs@swu.edu.cn}

\author{B. Saghai$^{2}$}\email{bijan.saghai@cea.fr}

\affiliation{1. School of Physical Science and Technology, Southwest University,
              Chongqing 400715, People's Republic of China\\
2. Universit\'e Paris-Saclay, Institut de Recherche sur les lois Fondamentales de l'Univers,
Irfu-CEA, F-91191 Gif-sur-Yvette, France}

\thispagestyle{empty}

\date{\today}

\begin{abstract}
In the extended chiral constituent quark model, the intrinsic $c \bar{c}$ content of the nucleon is investigated.
The probabilities of the quark-antiquark components in the nucleon wave functions are calculated by taking the nucleon
to be admixtures of three- and five-quark components, with the relevant transitions handled {\it via} the $^{3}$P$_{0}$
mechanism.
Predictions for the probability of  the $c \bar{c}$ in the nucleon wave function and the charmness-nucleon  sigma term
are presented.
Our numerical results turn out to be consistent with the predictions from various other approaches reported in the literature.
%
\end{abstract}
\pacs{12.39.-x, 12.38.Lg, 14.65.Bt, 14.65.Dw}
\maketitle

%
%
\section {Introduction}
Intrinsic quark-antiquark content of the nucleon is a prediction of Quantum Chromodynamics (QCD).
Recent review papers witness of the ongoing extensive theoretical and experimental efforts since about four decades on the light and heavy
quark-antiquark pairs in the baryons; see e.g.~\cite{Brodsky:2015fna,Hobbs:2013bia,Chang:2014jba}.

In 1980s, Brodsky and collaborators~\cite{Brodsky:1980pb,Brodsky:1981se} postulated the existence of the $|uudc\bar{c}\rangle$ components in the proton
in order to account for the large cross section of charmness production in the proton-proton collisions.
They coined the term "intrinsic" to be distinguished from the extrinsic contributions arising from gluon splitting in perturbative QCD.
The outcome of the developed light-cone formalism is known as the BHPS model and suggests a probability of
$\mathcal{P}^{c\bar{c}}_N \approx$ 1\%  for the intrinsic  $c \bar{c}$ (IC) component in the proton.
Following that pioneer work, various phenomenological approaches were developed to extract  $\mathcal{P}^{c\bar{c}}_N$ from data,
such as photon-gluon fusion~\cite{Hoffmann:1983ah,Harris:1995jx,Martin:2009iq}, meson cloud model~\cite{Steffens:1999hx} and global QCD analysis
of parton distribution ~\cite{Jimenez-Delgado:2014zga}, leading to $\mathcal{P}^{c\bar{c}}_N \approx$0.3-1(\%).
Pumplin and collaborators~\cite{Pumplin:2005yf} removed some simplification assumptions of the BHPS model and evaluated the sensitivity of the
 hard-scattering data to the IC, concluding that the corresponding  probability can range between zero and 3\%.
More recently Dulat and collaborators~\cite{Dulat:2013hea} analyzed the parton distribution function (PDF) of the proton based on the NNLO approximation
of perturbative QCD and included the combined H1 and ZEUS data~\cite{Abramowicz:1900rp}, reaching the conclusion that the PDF uncertainties are just as
large as the IC effects.
In summary, the nucleon's IC remains elusive, however several studies predict measurable effects of such possible components in the  ongoing and/ or forthcoming
experiments at the
 LHC~\cite{Dulat:2013hea,Brodsky:2012vg,Goncalves:2008sw,Kniehl:2012ti,Bednyakov:2013zta,Bailas:2015jlc,Ball:2015tna,Boettcher:2015sqn} and
 RHIC~\cite{Goncalves:2008sw,Kniehl:2009ar,Goncalves:2015apa}.

Another important entity in this realm is the charmness-nucleon sigma term $\sigma_{cN}$, related to the explicit breaking of chiral symmetry.
In recent years, few LQCD results became available by the ETM~\cite{Abdel-Rehim:2016won} and $\chi$QCD~\cite{Gong:2013vja} Collaborations.
It can also be extracted from another LQCD calculation performed by the MILC Collaboration~\cite{Freeman:2012ry}.
The central values coming from those works lie in the range of 67-94 MeV, albeit with large uncertainties $\approx$(30-50)\%,
making all results consistent with each other.

Phenomenologically, genuine higher Fock states in the baryons'  wave functions constitute a pertinent nonperturbative source of the intrinsic $Q \bar Q$
components.
In our recent works~\cite{An:2012kj,An:2014aea} we studied those components in baryons, with $Q \equiv u,~d,~s$, and the associated sigma terms.
In the present work we extend our approach to the intrinsic $c\bar{c}$  content of the nucleon and the charmness-nucleon sigma term.
Here we derive the wave functions for all possible quark-antiquark components in the nucleon, and calculate the corresponding probability amplitudes
using the $^{3}P_{0}$ quark-antiquark creation model~\cite{Le Yaouanc:1972ae}.
Also the resulting charmness-nucleon sigma term is evaluated.

The present manuscript is organized in the following way: in sec.~\ref{sec:Theo}, after a brief presentation of the theoretical frame, we give explicit
expressions for the sigma terms relating them to the quark-antiquark pair probabilities.
Numerical results for the probabilities of light, strange and charm quark-antiquark pairs in the nucleon, as well as $\sigma_{\pi N}$, $\sigma_{s N}$ and
$\sigma_{c N}$ are reported in sec.~\ref{sec:Res} and compared to findings from other sources.
Finally, sec.~\ref{sec:conclu} contains a summary and conclusions.

%
\section{Theoretical frame}
\label{sec:Theo}
The extended chiral constituent quark model on the light and strangeness components of baryons was developed in~\cite{An:2012kj},
and applied to the sigma terms of baryons in~\cite{An:2014aea}.
 So, here we briefly present the main content of the formalism and extend it to the  charm sector.
\subsection{The extended chiral constituent quark model}
\label{sec:recall}
In the extended chiral constituent quark model, wave function for the nucleon reads,
\begin{equation}
 |\psi\rangle_{N}=\frac{1}{\mathcal{\sqrt{N}}}{\Big[}|qqq\rangle+
 \sum_{i,n_{r},l}C_{in_{r}l}|qqq(Q \bar{Q}),i,n_{r},l\rangle {\Big ]}\,,
\label{wfn}
\end{equation}
where the first term is the conventional wave function for the nucleon with three constituent quarks ($q \equiv u,~d$) and the second term is a sum over
 all possible higher Fock components with a $Q \bar{Q}$ pair; $Q \bar{Q} \equiv u \bar{u},~d \bar{d},~s \bar{s},~c \bar{c}$.
Different possible orbital-flavor-spin-color configurations of the four-quark subsystems in the five-quark system,  numbered by $i$; $n_{r}$ and $l$,  denote
the inner radial and orbital quantum numbers, respectively, while $C_{in_{r}l}/\sqrt{\mathcal{N}}$ represents the probability amplitude for the corresponding
 five-quark component.
As discussed explicitly in~\cite{An:2012kj}, here we only need to consider the five-quark configurations with $n_{r}=0$ and $l=1$, consequently, there are
 17 different configurations which can be classified in four categories according to the orbital and spin wave functions of the four-quark subsystem;
 the corresponding configurations are listed in Table~\ref{caco}, using the shorthand notation for Young tableaux.
Note that, the charmness configurations with flavor symmetry $[31]_{F}^{1}$ cannot form Fock components of the nucleon.

%
\begin{table}[t]
\caption{\footnotesize Categories (2$^{nd}$ line) and configurations (lines 3-8) for five-quark
components.
\label{caco}}
%
\begin{tabular}{ccccccccccc}
\hline\hline
i & Category / Config. && i & Category / Config. && i & Category / Config. && i & Category / Config.  \\
  & I / $[31]_{X}[22]_{S}$ &&  & II / $[31]_{X}[31]_{S}$ &&  & III / $[4]_{X}[22]_{S}$ &&  & IV / $[4]_{X}[31]_{S}$  \\
\hline
 1 & $[31]_{X}[4]_{FS}[22]_{F}[22]_{S}$    &&  5 & $[31]_{X}[4]_{FS}[31]^1_{F}[31]_{S}$ &&
 11 & $[4]_{X}[31]_{FS}[211]_{F}[22]_{S}$ &&  14 &$[4]_{X}[31]_{FS}[211]_{F}[31]_{S}$ \\
 2 & $[31]_{X}[31]_{FS}[211]_{F}[22]_{S}$  && 6 & $[31]_{X}[4]_{FS}[31]^2_{F}[31]_{S}$ &&
 12 & $[4]_{X}[31]_{FS}[31]^1_{F}[22]_{S}$&&  15 & $[4]_{X}[31]_{FS}[22]_{F}[31]_{S}$ \\
 3 & $[31]_{X}[31]_{FS}[31]^1_{F}[22]_{S}$ && 7 & $[31]_{X}[31]_{FS}[211]_{F}[31]_{S}$ &&
 13 & $[4]_{X}[31]_{FS}[31]^2_{F}[22]_{S}$ &&  16 & $[4]_{X}[31]_{FS}[31]^1_{F}[31]_{S}$ \\
 4 & $[31]_{X}[31]_{FS}[31]^2_{F}[22]_{S}$ && 8 & $[31]_{X}[31]_{FS}[22]_{F}[31]_{S}$ && & &&  17 & $[4]_{X}[31]_{FS}[31]^2_{F}[31]_{S}$ \\
   &                                       && 9 & $[31]_{X}[31]_{FS}[31]^1_{F}[31]_{S}$ && & && & \\
   &                                       && 10 & $[31]_{X}[31]_{FS}[31]^2_{F}[31]_{S}$ && & && & \\
\hline
\hline
\end{tabular}
\end{table}

In Table~\ref{wfc}, we construct the explicit wave functions of the studied flavor configurations of the four-quark subsystem in the charmness components in
the  nucleon.
%
\begin{table}[b]
\caption{\footnotesize Flavor wave functions of the charmness configurations studied here.
 Note that, the full wave functions are obtained by multiplying each column by the corresponding normalization factor.
\label{wfc}}
\renewcommand\tabcolsep{0.4cm}
\renewcommand{\arraystretch}{1}
\scriptsize
\begin{tabular}{ccccccccc}
\hline\hline
           &  $[22]_{F_1}$ & $[22]_{F_2}$ & $[31]^{2}_{F_1}$ & $[31]^{2}_{F_2}$ & $[31]^{2}_{F_3}$ &  $[211]_{F_1}$ & $[211]_{F_2}$ & $[211]_{F_3}$ \\
\hline \\
$uudc$     &   2   &  0   &  0   &  6   &   0   &   2   &  0    &   0    \\
$uucd$     &   2   &  0   &  2   &  2   &   0   &  -2   &  0    &   0    \\
$dcuu$     &   2   &  0   &  -1  & -4   &  -2   &   0   &  2    &   1    \\
$cduu$     &   2   &  0   &  -1  & -4   &   2   &   0   &  -2   &   -1    \\
$duuc$     &  -1   & -1   &   0  & -3   &  -3   &  -1   &  -3   &   0     \\
$uduc$     &  -1   &  1   &   0  & -3   &   3   &  -1   &  3    &   0      \\
$cudu$     &  -1   &  1   &  -1  & 5    &   -1  &  -1   &  -1   &   1      \\
$ucdu$     &  -1   & -1   &  -1  &  5   &   1   &  -1   &   1   &   -1     \\
$cuud$     &  -1   & -1   &   2  &  -1  &   -1  &  1    &   3   &    0      \\
$ducu$     &  -1   &  1   &  -1  &  -1  &   -3   &  1   &   1   &   -1       \\
$ucud$     &  -1   &  1   &   2  &  -1  &   1    &  1   &  -3   &   0          \\
$udcu$     &  -1   & -1   &  -1  &  -1   &  3    &   1   &  -1  &   1         \\

\hline
\hline
\end{tabular}
\end{table}
%
To estimate the energies of the studied hidden charm pentaquark configurations, we employ the chiral constituent quark model developed
in~\cite{An:2012kj}.
 As discussed in that reference, all the configurations share a same energy $E_{0}=2127$~MeV, if the difference between the constituent masses of charm
and light quarks and the hyperfine interaction between quarks are not taken into account.
Consequently, the energy $E_{i}$ for the $i^{th}$ configuration reads,
\begin{equation}
E_{i}=E_{0}+2\delta m+E_{i}^{h}\, ,
\end{equation}
where $\delta m=m_c - m_q$ is the constituent mass difference between charm and light quarks, and $E_{i}^{h}$ the energy caused by hyperfine interaction
 between quarks.
To consider the hyperfine interaction between quarks, we employ the flavor-spin dependent version in the chiral constituent quark model~\cite{Glozman:1995fu},
\begin{eqnarray}
 H_{h}&=&-\sum_{i<j}\vec{\sigma}_{i}\cdot\vec{\sigma}_{j}
                    \Big [ \sum_{a=1}^{3}V_{\pi}(\vec{r}_{ij})\lambda^{a}_{i}\lambda^{a}_{j}+
                   \sum_{a=4}^{7}V_{K}(\vec{r}_{ij})\lambda^{a}_{i}\lambda^{a}_{j}+
                   V_{\eta}(\vec{r}_{ij})\lambda^{8}_{i}\lambda^{8}_{j}   \nonumber\\
                 && +\sum_{a=9}^{12}V_{D}(\vec{r}_{ij})\lambda^{a}_{i}\lambda^{a}_{j}
                 +\sum_{a=13}^{14}V_{D_{s}}(\vec{r}_{ij})\lambda^{a}_{i}\lambda^{a}_{j}
                 +V_{\eta_{c}}(\vec{r}_{ij})\lambda^{15}_{i}\lambda^{15}_{j}                    
                 \Big ]\,,
\label{hyp}
\end{eqnarray}
where $\lambda^{a}_{i}$ denotes the $SU(4)$ Gell-Mann matrix acting on the $i^{th}$ quark, $V_{M}(r_{ij})$ is the potential
of the $M$ meson-exchange interaction between $i^{th}$ and $j^{th}$ quark, as extensively discussed
in~\cite{Glozman:1995fu,Glozman:1995xy}.
Then, $E_{i}^{h}$ is obtained by
\begin{eqnarray}
E^{h}_{i}&=&\langle QQQ(Q\bar{Q}),i,0,1|H_{h}|QQQ(Q\bar{Q}),i,0,1\rangle  \nonumber\\
&=&-6 \sum_{njklm}
\Bigg [
(C_{[31]^{n}_{i}[211]_{n}}^{[1^4]})^{2}C^{[31]_{i}^{n}}_{[\mathcal{FS}]^{j}_{i}
[\mathcal{X}]_{i}^{l}}
C^{[31]_{i}^{n}}_{[\mathcal{FS}]^{k}_{i}[\mathcal{X}]_{i}^{m}}
\nonumber\\
&&
\Big (
\langle[\mathcal{X}]^{l}_{i}|V_{\pi}(r_{12})|[\mathcal{X}]^{m}_{i}\rangle
\langle[\mathcal{FS}]^{j}_{i}|\vec{\sigma}_{1}\cdot\vec{\sigma}_{2}\sum_{a=1}^{3}\lambda^{a}_{1}
\lambda^{a}_{2}|[\mathcal{FS}]^{k}_{i}\rangle
\nonumber\\
&& +\langle[\mathcal{X}]^{l}_{i}|V_{K}(\vec{r}_{12})|[\mathcal{X}]^{m}_{i}\rangle
\langle[\mathcal{FS}]^{j}_{i}|\vec{\sigma}_{1}\cdot\vec{\sigma}_{2}\sum_{a=4}^{7}\lambda^{a}_{1}
\lambda^{a}_{2}|[\mathcal{FS}]^{k}_{i}\rangle
\nonumber\\
&& +\langle[\mathcal{X}]^{l}_{i}|V_{\eta}(\vec{r}_{12})|[\mathcal{X}]^{m}_{i}\rangle
\langle[\mathcal{FS}]^{j}_{i}|\vec{\sigma}_{1}\cdot\vec{\sigma}_{2}
\lambda^{8}_{1} \lambda^{8}_{2}|[\mathcal{FS}]^{k}_{i}\rangle
\nonumber\\
&& +\langle[\mathcal{X}]^{l}_{i}|V_{D}(\vec{r}_{12})|[\mathcal{X}]^{m}_{i}\rangle
\langle[\mathcal{FS}]^{j}_{i}|\vec{\sigma}_{1}\cdot\vec{\sigma}_{2}\sum_{a=9}^{12}\lambda^{a}_{1}
\lambda^{a}_{2}|[\mathcal{FS}]^{k}_{i}\rangle
\nonumber\\
&& +\langle[\mathcal{X}]^{l}_{i}|V_{D_s}(\vec{r}_{12})|[\mathcal{X}]^{m}_{i}\rangle
\langle[\mathcal{FS}]^{j}_{i}|\vec{\sigma}_{1}\cdot\vec{\sigma}_{2}\sum_{a=13}^{14}\lambda^{a}_{1}
\lambda^{a}_{2}|[\mathcal{FS}]^{k}_{i}\rangle
\nonumber\\
&&+\langle[\mathcal{X}]^{l}_{i}|V_{\eta_c}(\vec{r}_{12})|[\mathcal{X}]^{m}_{i}\rangle
\langle[\mathcal{FS}]^{j}_{i}|\vec{\sigma}_{1}\cdot\vec{\sigma}_{2}\lambda^{15}_{1}
\lambda^{15}_{2}|[\mathcal{FS}]^{k}_{i}\rangle
 \Big )
\Bigg ]
\,,
\end{eqnarray}
where  $[\mathcal{FS}]^{\it N}_{i}$ and $[\mathcal{X}]^{\it N}_{i}$  represent the ${\it N}^{th}$ flavor-spin and orbital wave
functions of the four-quark subsystem in the five-quark configuration with number $i$ of the 17 five-quark configurations.
$C_{[31]^{n}_{i}[211]_{n}}^{[1^4]}$, $C^{[31]_{i}^{n}}_{[\mathcal{FS}]^{j}_{i}[\mathcal{X}]_{i}^{l}}$
and $C^{[31]_{i}^{n}}_{[\mathcal{FS}]^{k}_{i}[\mathcal{X}]_{i}^{m}}$ are the $S_{4}$ Clebsch-Gordan coefficients.

On the other hand, we have to consider the $SU(4)$ flavor symmetry breaking effects because of the large difference between the light and charm quark constituent masses.
 As introduced in~\cite{Yuan:2012zs}, these effects can be calculated by using the following flavor-dependent Hamiltonian
\begin{equation}
H_{sb}=-\sum_{i=1}^{4}\frac{m_{c}-m_{q}}{2m_{q}}\{\frac{p_{i}^{2}}{m_{c}}
+\frac{p_{\bar{c}}^{2}}{m_{c}}\}\delta_{ic}\,
\end{equation}
where $\delta_{ic}$ is a flavor dependent operator with eigenvalue $1$ for charm quark and $0$ for light quark.
Here $m_{q}$ is the constituent mass of the light quark.

The coefficient $C_{in_{r}l}$ for a given five-quark component can be related to the transition matrix element between the  three- and five-quark configurations
of the studied baryon. To calculate the corresponding transition matrix element, we use a $^{3}$P$_{0}$ version for the transition coupling operator $\hat{T}$,
\begin{eqnarray}
 \hat{T}&=&-\gamma\sum_{j}\mathcal{F}_{j,5}^{00}\mathcal{C}_{j,5}^{00}C_{OFSC}
 \sum_{m} \langle1,m;1,-m|00\rangle\chi^{1,m}_{j,5}\nonumber \\
&&\mathcal{Y}^{1,-m}_{j,5} (\vec{p}_{j}-\vec{p}_{5})b^{\dag}(\vec{p}_{j})d^{\dag}(\vec{p}_{5})\,,
\label{op}
\end{eqnarray}
with $\gamma$ a dimensionless constant of the model, $\mathcal{F}_{i,5}^{00}$ and $\mathcal{C}_{i,5}^{00}$ the flavor and color singlet of the quark-antiquark pair
$Q_{i} \bar{Q}$ in the five-quark system, and $C_{OFSC}$ an operator to calculate the orbital-flavor-spin-color overlap between the residual three-quark configuration
in the five-quark system and the valence three-quark system.

The probability of the  quark-antiquark pairs in the nucleon and the normalization factor read, respectively,
\begin{eqnarray}
\mathcal{P}^{Q \bar{Q}}&=& \frac{1}{\mathcal{N}}
\sum_{i=1}^{17}\Bigg[ \Big ( \frac{T_i^{Q \bar{Q}} }{M_N-E_i^{Q \bar{Q}}} \Big )^2 \Bigg ], \\ [10pt]
\label{prob}
\mathcal{N}  &\equiv&  1+ \sum_{i=1}^{17} \mathcal{N}_i
=1+\sum_{i=1}^{17} \sum_{Q \bar{Q}} \Bigg[ \Big ( \frac{T_i^{Q \bar{Q}}}{M_B-E_i^{Q \bar{Q}}}\Big )^2  \Bigg ].
\label{norm}
\end{eqnarray}
where the first term in Eq.~(\ref{norm})~is due to the valence three-quark state, while the second term comes from the five-quark mixtures.

To derive the explicit wave functions of the five-quark components with light quark-antiquark pairs in the nucleon, we construct the
wave functions for $u\bar{u}$ and $d\bar{d}$ components of the 17 different flavor-spin configurations to form the isospin state
$|\frac{1}{2},\frac{1}{2}\rangle$.
The five-quark components with $s\bar{s}$ and $c\bar{c}$ pairs form the nucleon isospin naturally since those pairs do not contribute
to isospin.
Accordingly, starting from Eq.~(\ref{prob}), the light quark-antiquark pairs ($u \bar{u}$ and $d \bar{d}$) probabilities for the nucleon
in terms of the five-quark probabilities per configuration ($P_N(i)$, $i$=1-17) are combined~\cite{An:2014aea}  with the relevant squared
Clebsch-Gordan coefficients in $SU(2)$ isospin space.
For the $s \bar {s}$ and $c\bar{c}$ components, the probabilities $\mathcal{P}_{N}^{s \bar s}$ and
$\mathcal{P}_{N}^{c \bar c}$ are obtained by summing up linearly the relevant nonvanishing contributions, ${P}_{N}^{s \bar s} (i)$
and ${P}_{N}^{c \bar c} (i)$ (i=1,17), respectively.
%
%
\subsection{Sigma terms}
\label{sec:s-t}
%
%
Here we proceed in line with Ref.~\cite{An:2014aea}, where explicit expressions for the pion- and strangeness-baryon sigma terms
were given as a function of quark-antiquark pairs probabilities.
Accordingly, the charmness-nucleon $\sigma$ term is defined as follows:
\begin{eqnarray}
\sigma_{c_N} & = & m_{c} \langle N|c \bar{c}|N\rangle,
\end{eqnarray}
which can be related to $\sigma_{\pi N}$
\begin{eqnarray}
\sigma_{c N} &=& \frac{m_c}{m_l}  \frac{ \langle N|c \bar{c}|N\rangle}{\langle N|u \bar{u} +d \bar{d}|N\rangle} \sigma_{\pi N} \\ [10pt]
&=&\frac{m_c}{m_l} \frac{2\mathcal{P}_N^{c \bar c
  }}{3+2(\mathcal{P}_N^{u \bar u}+\mathcal{P}_N^{d \bar d})} \sigma_{\pi N}
\end{eqnarray}
where
\begin{eqnarray}
 \label{piNy}
\sigma_{\pi N} &=& \frac{\hat{\sigma}}{1- 2({\langle N|s \bar{s}|N\rangle} / {\langle N|u \bar{u} +d \bar{d}|N\rangle})} \\ [10pt]
&=&\frac{3+2(\mathcal{P}_N^{u \bar u}+\mathcal{P}_N^{d \bar d})}
     {3+2(\mathcal{P}_N^{u \bar u}+\mathcal{P}_N^{d \bar d}-2\mathcal{P}_N^{s \bar s})} \hat{\sigma},
\end{eqnarray}
with $m_{l}\equiv(m_{u}+m_{d})/2$ the average current mass of the up and down quarks; $m_{s}$ and $m_{c}$ the current mass of the
strange and charm  quarks, respectively, and $\hat{\sigma}$ the nucleon expectation value of the purely octet operator.
%
%
\section{Results and Discussion}
\label{sec:Res}
First we present the model parameters.
As documented in~\cite{An:2013daa}, the input parameters of the model for ${u \bar u}$, ${d \bar d}$ and ${s \bar s}$ are taken from
the literature.
The newly introduced ones concern the charm quark, namely, its mass, and the hyperfine interaction strength
 between the light and charm quarks.
 For the former one, we use the empirical value $m_{c}=1275(25)$~MeV given in PDG~\cite{Agashe:2014kda}, and the latter one
is taken from Ref.~\cite{Glozman:1995xy}.

The only source of uncertainty in the probabilities, presented in sec.~\ref{sec:ResR},  comes from a common factor of the matrix elements of the transitions
 between three- and five-quark components and was found~\cite{An:2012kj} to be $V$=570$\pm$46 MeV, by successfully fitting the experimental data
for the proton flavor asymmetry $\bar{d} - \bar{u} \equiv {\mathcal{P}_p^{d \bar d}} - {\mathcal{P}_p^{u \bar u}}=0.118\pm0.012$~\cite{Towell:2001nh}.
Introduction of the five-quark components with the  charm quark-antiquark pairs and fitting the same data point,
the new extracted value is $V$=572$\pm$47 MeV, differing by 0.4\%  from the previous one.
For the $\sigma$-terms two additional entities contribute to the uncertainties~\cite{An:2014aea}, namely, the nonsinglet component
$\hat{\sigma}=33(5)$~MeV, as extracted within the chiral perturbation theory~\cite{Borasoy:1996bx} and the PDG masses
ratio~\cite{Agashe:2014kda}  $m_{s}/m_{l} = 27.5 (1.0)$.
Accordingly, compared to our previous studies~\cite{An:2012kj,An:2014aea} only one parameter was slightly readjusted in
the frame of the present work.

In  this section, we report our numerical results for the probabilities of the quark-antiquark components  in the  nucleon and the
 relevant sigma terms,  followed by comparisons to findings by other authors.
%
%
\subsection{Numerical results}
\label{sec:ResR}
Table~\ref{proba} embodies our numerical results.
In columns 3 to 5 the  quark-antiquark pairs probabilities in the nucleon per configuration are given for light, strange and charm components, respectively.
The total five quark-antiquark probabilities are reported in column 6. The pion-, strangeness- and charmness-sigma terms are given in columns
7 to 9, respectively.

Note that,the  numerical results for the light and strange quark-antiquark pairs (columns 3-4 and 7--8) , reported in~\cite{An:2014aea} for
$V$=570$\pm$46 MeV, are given here with the updated value for $V$ and allow us  to make clear the  relative weight of the ${c \bar{c}}$ component  and
the discussion on the sigma terms at the end of this section.

As reported in Table~\ref{proba},  out of the 17 five-quark configurations in the nucleon, only 3 of them contribute to all the light, $s \bar {s}$  and $c\bar{c}$
pairs probabilities, whereas 5 of them have only $u \bar {u}$ and / or  $d \bar {d}$ components, while the remaining 9 configurations are exclusively
composed of $s \bar {s}$ and $c\bar{c}$ pairs.

In the light quark-antiquark sector, the five-quark probability is dominated by the first category, where the total spin of the four-quark subsystem is $0$.
Within that category, the configuration $n^\circ$~1 gives the largest contribution and corresponds to the configuration with the lowest energy and largest
coupling to the three-quark component.
%
\begin{table}[hbt]
\caption{\footnotesize Predictions for probabilities of different five-quark
configurations for the nucleon (in \%), with
$\mathcal{P}^{q\bar{q}}_N = \mathcal{P}^{u\bar{u}}_N + \mathcal{P}^{d\bar{d}}_N$,
$\mathcal{P}^{Q\bar{Q}}_N = \mathcal{P}^{q\bar{q}}_N + \mathcal{P}^{s\bar{s}}_N+ \mathcal{P}^{c\bar{c}}_N$,
 and pion-, strangeness- and charmness-nucleon sigma terms (in MeV).
\label{proba}}
\scriptsize
\begin{tabular}{rccccccccc}
\hline\hline
i & Category  & $\mathcal{P}^{q\bar{q}}_N$ & $\mathcal{P}^{s\bar{s}}_N$ & $\mathcal{P}^{c\bar{c}}_{N}$ & $\mathcal{P}^{Q\bar{Q}}_N$  && $\sigma_{\pi N}$ & $\sigma_{sN}$ & $\sigma_{cN}$  \\
%
%
\hline \\
  & ~~~~I) $[31]_{X}[22]_{S}$: & &&  \\
 1&            &  14.58 (1.50)  &  0.98 (10)    & 0.04 (0)          & 15.60 (1.60)     &&  33.4 (5.1)     & 5.5 (1.5)   & 3.0 (8)    \\
 2&            &    0                 &  0.36 (4)     & 0.03 (1)           & 0.39 (4)            &&  33.2 (5.0)     & 2.2 (6)     & 2.7 (7)  \\
 3&            &   1.64 (17)      &   0              & 0                     & 1.64 (17)           &&  33.0 (5.0)     & 0        & 0     \\
 4&            &   0                  &  0.26 (3)      & 0.03 (1)          & 0.29 (3)            &&  33.1 (5.0)     & 1.6 (5)     & 2.6 (7) \\
  &~Category~I & 16.22 (1.66)  & 1.60 (16)  & 0.10 (1)      & 17.92 (1.83)     &&  33.6 (5.2)     & 8.9 (2.5)  & 7.8 (2.1) \\ \\

%
%
  & ~~~~II) $[31]_{X}[31]_{S}$: & &&\\
5 &            &  7.27 (75)      &  0              & 0                   & 7.27 (75)          &&  33.0 (5.0)     & 0       & 0 \\
6 &            &  0                  &  0.63 (6)      & 0.04 (0)       & 0.67 (6)           &&  33.3 (5.1)     & 3.8 (1.1)  & 3.2 (9)\\
7 &            &  0                 & 0.32 (4)       & 0.03 (1)        & 0.36 (3)           &&  33.1 (5.0)     & 2.0 (6)    & 2.6 (7)\\
8 &            &  0.61 (6)       & 0.18 (2)       & 0.02 (0)        & 0.81 (8)           &&  33.1 (5.0)     & 1.1 (3)    & 1.5 (4)\\
9 &            &  0.47 (5)       & 0                 & 0                   & 0.47 (5)           &&  33.0 (5.0)    & 0         & 0  \\
10&            &  0                 & 0.08 (1)       & 0.01 (0)        & 0.09 (1)          &&  33.0 (5.0)    & 0.5 (1)    & 0.8 (2) \\
  &~Category~II& 8.33 (0.86)  & 1.21 (13)  & 0.10 (1)     & 9.64 (98)        &&  33.5 (5.1)    & 7.0 (2.0)  & 7.8 (2.1) \\ \\
%
%
  & ~~~~III) $[4]_{X}[22]_{S}$: & &\\
11&            & 0                 & 0.85 (9)       & 0.09 (1)        & 0.94 (10)          &&  33.4 (5.1)    & 5.2 (1.5)  & 7.2 (2.0)\\
12&            & 4.13 (42)     & 0                & 0                   & 4.13 (42)          &&  33.0 (5.0)    & 0         & 0   \\
13&            & 0                 & 0.65 (7)     & 0.09(1)          & 0.74 (8)            &&  33.3 (5.1)     & 4.0(1.2)  & 7.0 (1.9)\\
& ~Category III& 4.13 (42) & 1.50 (16)   & 0.18 (2)       & 5.81 (60)         &&  33.7 (5.2)    & 9.0 (2.6) & 14.0 (3.8) \\ \\

%
  & ~~~~IV) $[4]_{X}[31]_{S}$: & &\\
14&            & 0                  & 0.77 (8)        & 0.09 (1)           & 0.86 (9)        &&  33.3 (5.1)   & 4.7 (1.4)    & 7.1 (2.0)\\
15&            & 1.49 (16)      & 0.44 (5)        & 0.06 (1)           & 1.99 (21)       &&  33.2 (5.0)   & 2.6 (8)      & 4.5 (1.2)\\
16&            & 1.18 (12)      & 0                  & 0                     & 1.18 (12)       &&  33.0 (5.0)   & 0           & 0 \\
17&            & 0                  & 0.19 (2)        & 0.03 (1)           & 0.22 (2)         &&  33.1 (5.1)   & 1.1 (3)      & 2.2 (6)\\
& ~Category~IV & 2.67 (28)  & 1.40 (15)  & 0.18 (2)           & 4.25 (44)       &&  33.6 (5.2)   & 8.5 (2.5)   & 13.8 (3.8) \\ \\
& All configurations & 31.35 (3.21) & 5.71 (59) & 0.56 (6)   & 37.62 (3.85)   &&  35.2 (5.5)   & 30.5 (8.5) & 39.3 (10.3)\\ \\
\hline
\hline
\end{tabular}
\end{table}

In the case of $\mathcal{P}^{s\bar{s}}_N$, the four categories have comparable contributions, though the first one gives the highest probability;
where the total spin of the four-quark subsystem is $S=1$ and the total angular momentum is $J=0$.

Finally, for $\mathcal{P}^{c\bar{c}}_{N}$, the last  two categories contribute almost equally, but with larger probabilities than the first two ones.
While the category III corresponds to  the four-quark subsystem $J=0$, in  the category IV the total spin of the four-quark
 subsystem should be $S_{[31]}=1$ and $J=S_{4}\oplus L_{\bar{q}}=0$.

 With respect to the sigma terms, the pion-nucleon $\sigma$ term of every configuration  is  $\approx$33.6~MeV, very close to the complete
 calculation with all 17 configurations leading to  $\approx$35~MeV.
But the strangeness- and charmness-nucleon sigma terms per configuration are about a factor of 4 to more than one order of magnitude
smaller than the total of all configurations.
Accordingly, any configuration truncated model will significantly underestimate both $\sigma_{sN}$ and $\sigma_{cN}$, leading to confusing results.
\subsection{Discussion and comparisons to previous results}
\label{sec:ResD}
Probabilities and sigma terms related to the light and strange quark-antiquark sector were presented and discussed in our previous
study~\cite{An:2014aea}. Therein, the determined probabilities came out compatible with those reported within  the generalized BHPS
approach~\cite{Chang:2011vx} and the  meson cloud model~\cite{arXiv:1002.4747}.
Also the  sigma terms related to the light and strange quark-antiquark sector turned out to be in good agreement with results
coming from various approaches, namely,  chiral Lagrangian~\cite{Semke:2012gs}, chiral perturbation theory~\cite{Ren:2013dzt},
and  LQCD~\cite{Durr:2011mp}.
 In this section we hence concentrate on the charm issues.
%
%
 \subsubsection{Quark-antiquark probabilities in the nucleon}
\label{sec:ResD-P}
  In Table~\ref{compaP} probabilities for the  $c \bar{c}$ pairs in the nucleon are reported.
%
%
\begin{table}[htb]
\begin{center}
\caption{\footnotesize Predictions for the probability of $c \bar{c}$ in the nucleon (\%).}
\label{compaP}
\renewcommand\tabcolsep{0.4cm}
\renewcommand{\arraystretch}{0.5}
\begin{tabular}{lcc}
\hline\hline
Reference                                                             &  Approach & $\mathcal{P}^{c\bar{c}}_{N}$  \\
\hline
Present work                                                          & E$\chi$CQM & 0.6(1)  \\
Brodsky {\it et al.}~\cite{Brodsky:1980pb}               & Light-cone  &  $\approx$1  \\
 Hoffmann and Moore,~\cite{Hoffmann:1983ah}     & PGF - NLO    &      0.31   \\
Harris {\it et al.}~\cite{Harris:1995jx}                        & PGF - NLO    &   0.86(60)   \\
Martin {\it et al.}~\cite{Martin:2009iq}                        & NNLO           &     0.3     \\
Steffens {\it et al.}~\cite{Steffens:1999hx}                & Meson cloud   &   $\approx$ 0.4       \\
Dulat {\it et al.}~\cite{Dulat:2013hea}                        &PQCD-NNLO    &  $\le$ 2     \\
Jimenez-Delgado~\cite{Jimenez-Delgado:2014zga} & PDF            &  0.3-0.4~;~ $\approx$1  \\
\hline
\hline
\end{tabular}
\end{center}
\end{table}

 As mentioned in Introduction, the first calculations embodying intrinsic $c \bar{c}$ was performed by Brodsky and collaborators
 ~\cite{Brodsky:1980pb} within the light-cone Fock space framework,  by introducing the hypothesis that ${P}^{c\bar{c}}_N$ could
 be around 1\%.
 Hoffmann and Moore,~\cite{Hoffmann:1983ah} investigated the matter within a photon-gluon fusion (PGF) model at NLO and also
 took into account the quark and target mass  contributions to the charm cross section, finding a smaller probability: ${P}^{c\bar{c}}_N$=0.31\%.
 Harris {\it et al.}~\cite{Harris:1995jx} extended that work interpreting the EMC charm production data~\cite{Aubert:1981ix} by calculating next-to-leading
order and generalizing it for both extrinsic and intrinsic contributions to the charm structure function and found an intrinsic charm probability of (0.86$\pm$0.60)\%.
 Steffens {\it et al.}~\cite{Steffens:1999hx} used a more extended data base for the charm structure function, including the very low-x region measurements by
the H1~\cite{Adloff:1996xq} and ZEUS~\cite{Breitweg:1997mj} collaborations.
The authors performed a consistent interpolation between the two asymptotic regions of massless evolution at large $Q^2$ and the PGF, finding a slight preference
for ${P}^{c\bar{c}}_N \approx$0.4\%.
Later Martin {\it et al.}~\cite{Martin:2009iq}, using very extensive data coming from some 40 data sets released between 1989 and 2008, updated the parton
distribution functions  determined from global analysis of hard-scattering data up to NNLO and found ${P}^{c\bar{c}}_N$=0.3\%.
In a  recent work, Jimenez-Delgado {\it et al.}~\cite{Jimenez-Delgado:2014zga} report the results of a new global QCD analysis of parton distribution functions (PDF),
concentrating on the momentum fraction carried by the intrinsic charm quarks in terms of the Feynman-x,
\begin{equation}
{\langle x \rangle}_{c + \bar{c}} = \int_{0}^{1} x {\Big [c(x) + \bar{c} (x) \Big ]}  dx,
\label{Mom}
\end{equation}
which is related to the $c  \bar{c}$ content of the nucleon by,
\begin{equation}
\mathcal{P}^{c\bar{c}}_{N} = \int_{0}^{1}  c(x)   dx =  \int_{0}^{1}  \bar{c}(x)   dx
\label{Nc}
\end{equation}

Note that in the BHPS model the predictions~\cite{Brodsky:2015fna} at the input scale $Q_0=m_c$=1.3 GeV,  are ${\langle x \rangle}_{c + \bar{c}}$ =0.57\% and
$\mathcal{P}^{c\bar{c}}_{N}$ = 1\%.
Jimenez-Delgado {\it et al.}~\cite{Jimenez-Delgado:2014zga} analyzed a large set of data with ${\langle x \rangle}_{c + \bar{c}}$ in the range of 0 to $\approx$0.6\%.
Fitting only the EMC data, they obtained $\mathcal{P}^{c\bar{c}}_{N}$ = 0.3-0.4\%, while excluding the old EMC data led to ${\langle x \rangle}_{c + \bar{c}}$ =0.5\%;
a value close to the BHPS prediction.
The minimization approach in the former work raised  a debate~\cite{Brodsky:2015uwa,Jimenez-Delgado:2015tma}
emphasizing the need for more precise data.

Then, a global conclusion on the probability of the intrinsic $c\bar{c}$ component in the nucleon is that its value would be in the range of 0.3 to 1\% and our result, 0.6\%,
falls in that range.

However, as briefly discussed below, the genuine ${c\bar{c}}$ component is predicted to play a significant role in the forthcoming measurements using high energy
beams at  CERN/LHC, BNL/RHIC...

Actually, it is well established that the heavy quarks produced in line with perturbative QCD carry small longitudinal momentum, while the intrinsic
heavy constituents transport the largest fraction of the momentum of the hadron.
Accordingly, to probe the intrinsic charm in the  nucleon various guidelines were elaborated.

Brodsky and collaborators~\cite{Brodsky:2012vg} proposed a fixed target experiment for the LHC 7 TeV beam, allowing precise enough measurements of the rapidity
distribution of open- or hidden-charm hadrons at $\sqrt{s_{NN}}$=115 GeV, accessing the domain of high $x_F$; knowing that~\cite{Brodsky:1980pb,Brodsky:2007yz}
the intrinsic quark-antiquark possible manifestations should be looked for roughly in the range 0.2$\leq x_F \leq$0.8.

Kniehl and collaborators~\cite{Kniehl:2012ti} employed the general-mass variable-flavor-number scheme at NLO to study the inclusive production of the $D$ meson,
$pp \to D^\circ  X$, at the LHC and found that the production cross sections at  $\sqrt{s_{NN}}$=7 TeV and  large values of rapidity are sensitive to a non-perturbative
component of the charm PDF for $\mathcal{P}^{c\bar{c}}_{N}$ = 3.5\% .

Bednyakov and collaborators~\cite{Bednyakov:2013zta} reported results for  $pp \to \gamma c X$ differential cross section  at $\sqrt{s}$=8 TeV.
Calculations come from the radiatively generated charm PDF (CTEQ66), the sea-like PDF  (CTEQ66c4) and the BHPS PDF (CTEQ66c2) also
 for $\mathcal{P}^{c\bar{c}}_{N}$ = 3.5\% and found that the IC manifestation could be measured with both the ATLAS and CMS detectors.

Bailas and Goncalves~\cite{Bailas:2015jlc} studied, within various models, the impact of the IC on the rapidity and transverse momentum distribution
in the Z-boson production in proton-proton collisions at the LHC, and showed that the $Z+c$ cross section is significantly sensitive to the presence
of the IC.

Finally, the relevance of the  CEBAF-12 GeV and FAIR-PANDA facilities to study the multiquark dynamics in baryons was also underlined~\cite{Yuan:2012wz}.
%
%
\subsubsection{Sigma terms}
\label{sec:ResD-S}
Compared to $\sigma_{\pi N}$ and, to a lesser extent to $\sigma_{sN}$, for the charmness-nucleon sigma term fewer results are available,
coming from lattice QCD results as given in Table~\ref{compaS}.
Note that, wherever appropriate, using statistical and systematic uncertainties reported in those papers, we give $\delta = \sqrt{\delta^2_{stat} + \delta^2_{sys}}$.

The most recent results were released by the ETM Collaboration~\cite{Abdel-Rehim:2016won}, employing improved methods for the disconnected quark loops;
determining $\sigma_{\pi N}$, $\sigma_{sN}$ and $\sigma_{cN}$.
 Comparing our results to those of the latter work, we note that the outcomes for $\sigma_{\pi N}$ and $\sigma_{sN}$ are in agreement within 1$\sigma$
 (for a comprehensive discussion with extractions of these latter terms by other authors see~\cite{An:2014aea} ).
 Our value for  $\sigma_{cN}$ is compatible with the ETM result within 2$\sigma$.
 This is also the case for the strangeness content of the nucleon $y_{_N}$, for which they get 0.075(16) compared to our approach's value 0.031(3).
 The same quantity for the charmless content of the nucleon within the present work is $y^c_{_N}$=0.004(1), but to our knowledge no other value
 was reported  in the literature for that entity.
 %
%
\begin{table}[htb]
\begin{center}
\caption{\footnotesize Predictions for the sigma terms $\sigma_{\pi N}$, $\sigma_{sN}$ and $\sigma_{cN}$ of the nucleon (MeV).}
\label{compaS}
\renewcommand\tabcolsep{0.4cm}
\renewcommand{\arraystretch}{0.5}
\begin{tabular}{lccccc}
\hline\hline
 Reference (Collaboration)&  Approach
 & ${\sigma}_{\pi N}$ & ${\sigma}_{s N}$& $\sigma_{cN}$
\\
\hline
Present work                                                                          & E$\chi$CQM & 35(6) &   30(8)   & 39(10) \\
Abdel-Rehim {\it et al.}~\cite{Abdel-Rehim:2016won} (ETM) & LQCD           &  37(7) &   41(8)  &  79(22)  \\
Gong {\it et al.}~\cite{Gong:2013vja} ($\chi$QCD)                & LQCD            &            &   33(6)   & 94(31) \\
Freeman and Toussaint~\cite{Freeman:2012ry} (MILC)        & LQCD            &            &   39(8)   & 67(32) \\
\hline
\hline
\end{tabular}
\end{center}
\end{table}
%

 The $\chi$QCD Collaboration~\cite{Gong:2013vja} investigated  the $Q \bar{Q}$  components within a dynamical LQCD with overlap valence quarks
 on 2+1 flavors domain-wall fermion gauge configuration. They performed calculations for the strange and charm quark-antiquark
 contributions and determined both $\sigma_{sN}$ and $\sigma_{cN}$.
 For the strangeness sigma term the agreement between their result and ours is perfect, while for $\sigma_{cN}$ the two finding are compatible with each other
 within less than 2$\sigma$.

 %
 The MILC Collaboration ~\cite{Freeman:2012ry} applied a hybrid method to the large library of improved staggered gauge configuration to calculate both
 matrix-elements $ \langle  N |s  \bar{s}|N\rangle$ = 0.44$\pm$0.08 (stat)$\pm$0.05 and $ \langle N|c \bar{c}|N\rangle$ = 0.058$\pm$0.027 (stat).
Using for the masses the values quoted  by the authors, $m_s$=89.0 MeV and $m_c$=1.2 GeV, we report in Table~\ref{compaS} the corresponding sigma terms.
Here also we find  good agreement with our results for $\sigma_{sN}$ and $\sigma_{cN}$ within 1$\sigma$.
%
%
\section {Summary and conclusions}
\label{sec:conclu}
Our recent  works~\cite{An:2012kj,An:2014aea} and the present study, performed within the extended chiral quark approach,  constitute a thorough
investigation of the genuine quark-antiquarks pairs in the nucleon.
The quark-antiquark pairs creation was calculated {\it via} the $^{3}P_{0}$ mechanism~\cite{Le Yaouanc:1972ae}.
All possible five-quark configurations which may form higher Fock components of the nucleon were taken into account and it was shown that any  configuration
 truncated calculation will lead to unrealistic results.

This coherent and comprehensive set of results allowed us to predict the probabilities of the $u \bar{u}$, $d \bar{d}$, $s \bar{s}$ and $c \bar{c}$
pairs in the nucleon as well as the associated sigma terms $\sigma_{\pi N}$, $\sigma_{sN}$ and $\sigma_{cN}$.
The model uncertainties are about 10\%, mainly due to the only fitted parameter on the proton flavor asymmetry
$\bar{d} - \bar{u}$=0.118(12)~\cite{Towell:2001nh}.
All other parameters were taken from the literature~\cite{An:2013daa}.

Extensive comparisons with the outcomes of other approaches, reported in \cite{An:2012kj,An:2014aea} and the present paper led, in general, to
compatibility of the obtained results with those found in the literature.
To our knowledge,  the present approach is the only available one putting forward predictions for all the above
mentioned entities within a single approach and set of input parameters.

The predicted probabilities of the five-quark components with light, strange and charm quark-antiquark pairs in the nucleon wave function,
turned out to be (in \%)  $P^{q\bar{q}}_N$=31.3(3.2), $P^{s\bar{s}}_N$=5.7 (6) and $P^{c\bar{c}}_{N}$=0.6(1), respectively,
adding up to $P^{Q\bar{Q}}_N$=37.6 (3.8).
As reported in ~\cite{An:2012kj} and Sec.~\ref{sec:ResD-P}, our findings are compatible with results released by several authors.

Here, three observations are in order: i) the intrinsic five-quark states represent a significant part of the nucleon wave function,
ii) the probability of charm-anticharm pairs is rather tiny, iii) there is no non-ambiguous experimental evidence for the existence of heavy quark-antiquark pairs in the
nonperturbative regime.
However, as discussed in Sec.~\ref{sec:ResD-P},  the present state-of-the-art in experimental high-energy physics allows us anticipating crucial
measurements at  the LHC and RHIC~\cite{Brodsky:2012vg,Goncalves:2008sw,Kniehl:2012ti,Bednyakov:2013zta,Bailas:2015jlc,Ball:2015tna,Boettcher:2015sqn,
Kniehl:2009ar,Goncalves:2015apa}.
Also empirical determination of the intrinsic charm through  PDF analysis with heavy-quarks are foreseen to shed a valuable light on those
issues; see e.g.~\cite{Ball:2015dpa,Butterworth:2015oua} and references therein.

 In parallel, LQCD calculations are producing results for the charmless-nucleon sigma term~\cite{Abdel-Rehim:2016won,Gong:2013vja,Freeman:2012ry}.
 Here also our determination of that entity is  compatible with the LQCD findings.
 Refinements in the latter approach, expected to reduce the presently large uncertainties, will certainly offer a better understanding of the underlying mechanisms
 with respect to the role, if any, played by the  charmness in the nucleon.

 Those efforts will hopefully lead to uncovering the puzzle of  possible charm  components in the nucleon.

%
%
%
\begin{acknowledgments}
We thank Y. Chen and L. S. Geng for the suggestion to consider the charmness-nucleon sigma term and useful discussions.
This work is partly supported by the Chongqing Natural Science Foundation under Grant No. cstc2015jcyjA00032, and the
Fundamental Research Funds for the Central Universities under Grants No. XDJK2015C150 and No. SWU115020.
\end{acknowledgments}

\begin{thebibliography}{99}
%
\bibitem{Brodsky:2015fna}
  S.~J.~Brodsky, A.~Kusina, F.~Lyonnet, I.~Schienbein, H.~Spiesberger and R.~Vogt,
  A review of the intrinsic heavy quark content of the nucleon,
  Adv.\ High Energy Phys.\  {\bf 2015}, 231547 (2015).

\bibitem{Hobbs:2013bia}
  T.~J.~Hobbs, J.~T.~Londergan and W.~Melnitchouk,
  Phenomenology of nonperturbative charm in the nucleon,
  Phys.\ Rev.\ D {\bf 89},  074008 (2014).

\bibitem{Chang:2014jba}
  W.~C.~Chang and J.~C.~Peng,
  Flavor Structure of the Nucleon Sea,
  Prog.\ Part.\ Nucl.\ Phys.\  {\bf 79}, 95 (2014).

\bibitem{Brodsky:1980pb}
  S.~J.~Brodsky, P.~Hoyer, C.~Peterson and N.~Sakai,
  The Intrinsic Charm of the Proton,
  Phys.\ Lett.\ B {\bf 93}, 451 (1980).

\bibitem{Brodsky:1981se}
  S.~J.~Brodsky, C.~Peterson and N.~Sakai,
  Intrinsic Heavy Quark States,
  Phys.\ Rev.\ D {\bf 23}, 2745 (1981).
%

\bibitem{Hoffmann:1983ah}
  E.~Hoffmann and R.~Moore,
  Subleading Contributions to the Intrinsic Charm of the Nucleon,
  Z.\ Phys.\ C {\bf 20}, 71 (1983).

\bibitem{Harris:1995jx}
  B.~W.~Harris, J.~Smith and R.~Vogt,
  Reanalysis of the EMC charm production data with extrinsic and intrinsic charm at NLO,
  Nucl.\ Phys.\ B {\bf 461}, 181 (1996).

\bibitem{Martin:2009iq}
  A.~D.~Martin, W.~J.~Stirling, R.~S.~Thorne and G.~Watt,
  Parton distributions for the LHC,
  Eur.\ Phys.\ J.\ C {\bf 63}, 189 (2009).

\bibitem{Steffens:1999hx}
  F.~M.~Steffens, W.~Melnitchouk and A.~W.~Thomas,
  Charm in the nucleon,
  Eur.\ Phys.\ J.\ C {\bf 11}, 673 (1999).

\bibitem{Jimenez-Delgado:2014zga}
  P.~Jimenez-Delgado, T.~J.~Hobbs, J.~T.~Londergan and W.~Melnitchouk,
  New limits on intrinsic charm in the nucleon from global analysis of parton distributions,
  Phys.\ Rev.\ Lett.\  {\bf 114},  082002 (2015).

\bibitem{Pumplin:2005yf}
  J.~Pumplin,
  Light-cone models for intrinsic charm and bottom,
  Phys.\ Rev.\ D {\bf 73}, 114015 (2006).
%
  J.~Pumplin, H.~L.~Lai and W.~K.~Tung,
  The Charm Parton Content of the Nucleon,
  Phys.\ Rev.\ D  {\bf 75}, 054029 (2007).

\bibitem{Dulat:2013hea}
  S.~Dulat  {\it et al.},
  Intrinsic Charm Parton Distribution Functions from CTEQ-TEA Global Analysis,
  Phys.\ Rev.\ D {\bf 89},  073004 (2014).

\bibitem{Abramowicz:1900rp}
  H.~Abramowicz {\it et al.} [H1 and ZEUS Collaborations],
  Combination and QCD Analysis of Charm Production Cross Section Measurements in Deep-Inelastic ep Scattering at HERA,
  Eur.\ Phys.\ J.\ C {\bf 73}, no. 2, 2311 (2013).

\bibitem{Brodsky:2012vg}
  S.~J.~Brodsky, F.~Fleuret, C.~Hadjidakis and J.~P.~Lansberg,
  Physics Opportunities of a Fixed-Target Experiment using the LHC Beams,
  Phys.\ Rept.\  {\bf 522}, 239 (2013).

\bibitem{Goncalves:2008sw}
  V.~P.~Goncalves and F.~S.~Navarra,
  Looking for intrinsic charm in the forward region at BNL RHIC and CERN LHC,
  Nucl.\ Phys.\ A {\bf 842}, 59 (2010).

\bibitem{Kniehl:2012ti}
  B.~A.~Kniehl, G.~Kramer, I.~Schienbein and H.~Spiesberger,
  Inclusive Charmed-Meson Production at the CERN LHC,
  Eur.\ Phys.\ J.\ C {\bf 72}, 2082 (2012).

\bibitem{Bednyakov:2013zta}
  V.~A.~Bednyakov  {\it et al.},
  Searching for intrinsic charm in the proton at the LHC,
  Phys.\ Lett.\ B {\bf 728}, 602 (2014).

\bibitem{Bailas:2015jlc}
  G.~Bailas and V.~P.~Goncalves,
  Phenomenological implications of the intrinsic charm in the $Z$ boson production at the LHC,
  Eur.\ Phys.\ J.\ C {\bf 76}, 105 (2016).

\bibitem{Ball:2015tna}
  R.~D.~Ball, V.~Bertone, M.~Bonvini, S.~Forte, P.~Groth Merrild, J.~Rojo and L.~Rottoli,
  Intrinsic charm in a matched general-mass scheme,
  Phys.\ Lett.\ B {\bf 754}, 49 (2016).

\bibitem{Boettcher:2015sqn}
  T.~Boettcher, P.~Ilten and M.~Williams,
  A direct probe of the intrinsic charm content of the proton,
  arXiv:1512.06666 [hep-ph].

\bibitem{Kniehl:2009ar}
  B.~A.~Kniehl, G.~Kramer, I.~Schienbein and H.~Spiesberger,
 Open charm hadroproduction and the charm content of the proton,
  Phys.\ Rev.\ D {\bf 79}, 094009 (2009).

\bibitem{Goncalves:2015apa}
  V.~P.~Gonçalves,
  Theoretical aspects of heavy-flavour production at ultra-high cosmic ray energies,
  EPJ Web Conf.\  {\bf 99}, 07001 (2015).
%

\bibitem{Abdel-Rehim:2016won}
  A.~Abdel-Rehim {\it et al.} [ETM Collaboration],
  Direct Evaluation of the Quark Content of the Nucleon from Lattice QCD at the Physical Point,
  arXiv:1601.01624 [hep-lat].

\bibitem{Gong:2013vja}
  M.~Gong {\it et al.} [XQCD Collaboration],
  Strangeness and charmness content of the nucleon from overlap fermions on 2+1-flavor domain-wall fermion configurations,
  Phys.\ Rev.\ D {\bf 88}, 014503 (2013).

\bibitem{Freeman:2012ry}
  W.~Freeman and D. Toussaint [MILC Collaboration],
  Intrinsic strangeness and charm of the nucleon using improved staggered fermions,
  Phys.\ Rev.\ D {\bf 88}, 054503 (2013).


\bibitem{An:2012kj}
  C.~S.~An and B.~Saghai,
  Sea flavor content of octet baryons and intrinsic five-quark Fock states,
  Phys.\ Rev.\ C {\bf 85}, 055203 (2012).

\bibitem{An:2014aea}
  C.~S.~An and B.~Saghai,
  Pion- and strangeness-baryon $\sigma$ terms in the extended chiral constituent quark model,
  Phys.\ Rev.\ D {\bf 92}, 014002 (2015).

\bibitem{Le Yaouanc:1972ae}
  A.~Le Yaouanc  {\it et al.},
  Naive quark pair creation model of strong interaction vertices,
  Phys.\ Rev.\ D {\bf 8}, 2223 (1973) ;
%
  A.~Le Yaouanc  {\it et al.},
  Naive quark pair creation model and baryon decays,
 Phys.\ Rev.\ D  {\bf 9}, 1415 (1974) ;
  R.~Kokoski and N.~Isgur,
  Meson Decays by Flux Tube Breaking,
   Phys.\ Rev.\ D  {\bf 35}, 907 (1987).
%
%

\bibitem{Glozman:1995fu}
  L.~Y.~Glozman and D.~O.~Riska,
  The Spectrum of the nucleons and the strange hyperons and chiral dynamics,
  Phys.\ Rept.\  {\bf 268}, 263 (1996).

\bibitem{Glozman:1995xy}
  L.~Y.~Glozman and D.~O.~Riska,
  The Charm and bottom hyperons and chiral dynamics,
  Nucl.\ Phys.\ A {\bf 603}, 326 (1996); Erratum: [Nucl.\ Phys.\ A {\bf 620}, 510 (1997)].

\bibitem{Yuan:2012zs}
  S.~G.~Yuan, C.~S.~An, K.~W.~Wei, B.~S.~Zou and H.~S.~Xu,
  Spectrum of low-lying $s^{3}Q\bar{Q}$ configurations with negative parity,
  Phys.\ Rev.\ C {\bf 87}, 025205 (2013).

\bibitem{An:2013daa}
  C.~S.~An and B.~Saghai,
  Strangeness magnetic form factor of the proton in the extended chiral quark model,
  Phys.\ Rev.\ C {\bf 88}, 025206 (2013).

%

\bibitem{Agashe:2014kda}
  K.~A.~Olive {\it et al.} [Particle Data Group Collaboration],
  Review of Particle Physics,
  Chin.\ Phys.\ C {\bf 38}, 090001 (2014).

\bibitem{Towell:2001nh}
  R.~S.~Towell {\it et al.} [NuSea Collaboration],
Improved measurement of the $\bar{d} $-$\bar{u}$ asymmetry in the nucleon sea,
  Phys.\ Rev.\ D {\bf 64}, 052002 (2001).

\bibitem{Borasoy:1996bx}
  B.~Borasoy and U.~-G.~Meissner,
  Chiral expansion of baryon masses and sigma terms,
  Annals Phys.\  {\bf 254}, 192 (1997).

\bibitem{Chang:2011vx}
  W.~C.~Chang and J.~C.~Peng,
  Flavor Asymmetry of the Nucleon Sea and the Five-Quark Components of the Nucleons,
  Phys.\ Rev.\ Lett.\  {\bf 106}, 252002 (2011);
%
  W.~C.~Chang and J.~C.~Peng,
  Extraction of Various Five-Quark Components of the Nucleons,
  Phys.\ Lett.\ B {\bf 704}, 197 (2011).

\bibitem{arXiv:1002.4747}
  L.~Shao, Y.~-J.~Zhang and B.~-Q.~Ma,
  Sea quark contents of octet baryons,
  Phys.\ Lett.\ B\ {\bf 686}, 136  (2010).

%
\bibitem{Semke:2012gs}
  A.~Semke and M.~F.~M.~Lutz,
  Strangeness in the baryon ground states,
  Phys.\ Lett.\ B {\bf 717}, 242 (2012);
%
  M.~F.~M.~Lutz {\it et al.},
  Finite volume effects in the chiral extrapolation of baryon masses,
  Phys.\ Rev.\ D {\bf 90}, 054505 (2014).
\bibitem{Ren:2013dzt}
  X.~L.~Ren {\it et al.},
  Virtual decuplet effects on octet baryon masses in covariant baryon chiral perturbation theory,
  Phys.\ Rev.\ D {\bf 87}, 074001 (2013);
%
  X.~L.~Ren {\it et al.},
  Scalar strangeness content of the nucleon and baryon sigma terms,
    Phys.\ Rev.\ D  {\bf 91}, 051502 (2015).
%
%
%
\bibitem{Durr:2011mp}
  S.~Durr {\it et al.},
  Sigma term and strangeness content of octet baryons,
  Phys.\ Rev.\ D {\bf 85}, 014509 (2012);
%
  R.~Horsley {\it et al.}  (QCDSF-UKQCD Collaboration),
  Hyperon sigma terms for 2+1 quark flavours,
  Phys.\ Rev.\ D {\bf 85}, 034506 (2012);
  G.~S.~Bali {\it et al.}  (QCDSF Collaboration),
  The strange and light quark contributions to the nucleon mass from Lattice QCD,
  Phys.\ Rev.\ D {\bf 85}, 054502 (2012);
%
  C.~Alexandrou {\it et al.},
  The quark contents of the nucleon and their implication for dark matter search,
  PoS LATTICE {\bf 2013}, 295 (2014).

\bibitem{Aubert:1981ix}
  J.~J.~Aubert {\it et al.} [European Muon Collaboration],
  An Experimental Limit on the Intrinsic Charm Component of the Nucleon,
  Phys.\ Lett.\ B {\bf 110}, 73 (1982);
%
  J.~J.~Aubert {\it et al.} [European Muon Collaboration],
  Production of charmed particles in 250-GeV $\mu^+$ - iron interactions,
  Nucl.\ Phys.\ B {\bf 213}, 31 (1983).

\bibitem{Adloff:1996xq}
  C.~Adloff {\it et al.} [H1 Collaboration],
  Inclusive D0 and D*+- production in deep inelastic e p scattering at HERA,
  Z.\ Phys.\ C {\bf 72}, 593 (1996).

\bibitem{Breitweg:1997mj}
  J.~Breitweg {\it et al.} [ZEUS Collaboration],
  D* production in deep inelastic scattering at HERA,
  Phys.\ Lett.\ B {\bf 407}, 402 (1997).

\bibitem{Brodsky:2015uwa}
  S.~J.~Brodsky and S.~Gardner,
  Comment on "New Limits on Intrinsic Charm in the Nucleon from Global Analysis of Parton Distributions",
  Phys.\ Rev.\ Lett.\  {\bf 116},   019101 (2016).

\bibitem{Jimenez-Delgado:2015tma}
  P.~Jimenez-Delgado, T.~J.~Hobbs, J.~T.~Londergan and W.~Melnitchouk,
  Reply to Comment on "New limits on intrinsic charm in the nucleon from global analysis of parton distributions",
  Phys.\ Rev.\ Lett.\  {\bf 116},  019102 (2016).

\bibitem{Brodsky:2007yz}
  S.~J.~Brodsky, A.~S.~Goldhaber, B.~Z.~Kopeliovich and I.~Schmidt,
  Higgs Hadroproduction at Large Feynman x,
  Nucl.\ Phys.\ B {\bf 807}, 334 (2009).

\bibitem{Yuan:2012wz}
  S.~G.~Yuan, K.~W.~Wei, J.~He, H.~S.~Xu and B.~S.~Zou,
  Study of $qqqc\bar{c}$ five quark system with three kinds of quark-quark hyperfine interaction,
  Eur.\ Phys.\ J.\ A {\bf 48}, 61 (2012);
  B.~S.~Zou,
  Hadron spectroscopy from strangeness to charm and beauty,
  Nucl.\ Phys.\ A {\bf 914}, 454 (2013).

\bibitem{Ball:2015dpa}
  R.~D.~Ball, M.~Bonvini and L.~Rottoli,
  Charm in Deep-Inelastic Scattering,
  JHEP {\bf 1511}, 122 (2015).

\bibitem{Butterworth:2015oua}
  J.~Butterworth {\it et al.},
  PDF4LHC recommendations for LHC Run II,
  J.\ Phys.\ G {\bf 43}, 023001 (2016).

\end{thebibliography}
\end{document}